\documentclass[runningheads]{svmult}

\usepackage{makeidx}   
\usepackage{graphicx}  
\usepackage{subeqnar}  
\usepackage{multicol}  
\usepackage{physprbb}  
\makeindex             

\usepackage{amssymb}
\newcommand{\Msun}{\rm{M}_{\odot}}

\newcommand{\Ex}[1]{\times10^{#1}}
\begin{document}
\title*{Massive Star Evolution Through the Ages}
\toctitle{Massive Star Evolution Through the Ages}
%
%
\titlerunning{Massive Star Evolution through the Ages}
%
\author{Alexander Heger\inst{1}
\and S.~E.~Woosley\inst{2}
\and C.~L.~Fryer\inst{3}
\and Norbert Langer\inst{4}
}
\authorrunning{Heger, Woosley, Fryer, \& Langer}
%
%
\institute{
Department of Astronomy and Astrophysics, 
Enrico Fermi Institute,\\
The University of Chicago,
5640 S.\ Ellis Ave,
Chicago, IL 60637,
U.S.A.
\and
Department of Astronomy and Astrophysics,
University of California, \\
Santa Cruz, CA 95064, U.S.A.
\and 
Theoretical Astrophysics, MS B288,
Los Alamos National Laboratories,\\ 
Los Alamos, NM 87545,
U.S.A.
\and
Astronomical Institute, 
P.O. Box 80000,
NL-3508 TA Utrecht, 
The Netherlands
}

\maketitle              

\begin{abstract}
We review the current basic picture of the evolution of massive stars
and how their evolution and structure changes as a function of initial
mass.  We give an overview of the fate of modern (Pop I) and
primordial (Pop III) stars with emphasis on massive and very massive
stars.  For single stars we show how the type of explosions, the type
of remnant and their frequencies changes for different initial
metallicities.
\end{abstract}

\section{Massive star evolution}

\begin{figure}[t]
\begin{center}
\includegraphics[angle=-90,width=\columnwidth,bb=30 21 580 741]{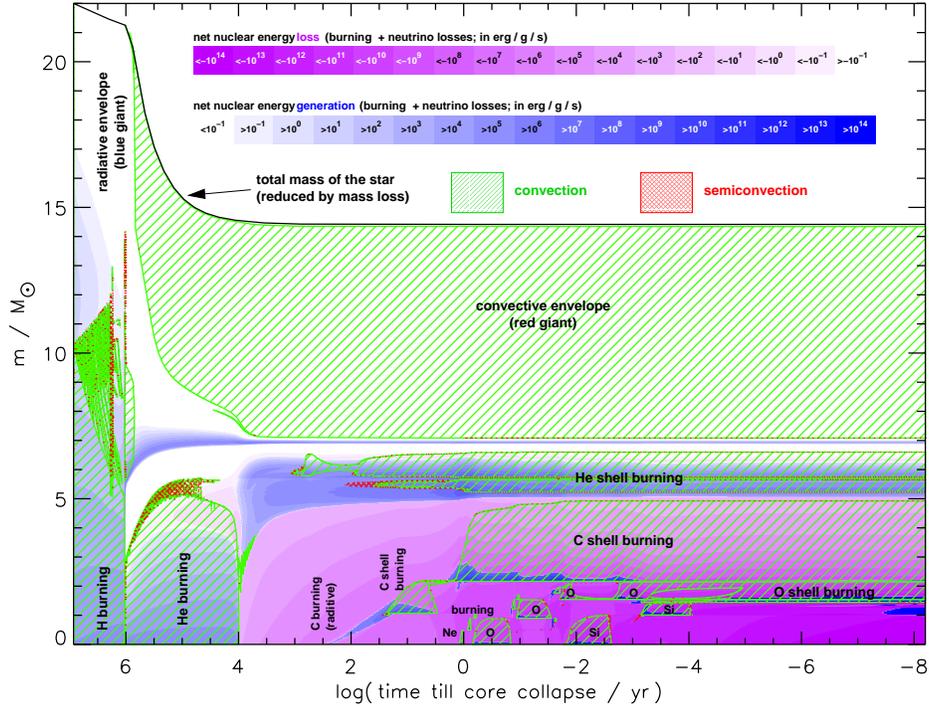}
\end{center}
\caption[]{Interior structure of a $22\,\Msun$ star of solar
composition as a function of time (logarithm of time till core
collapse) and enclosed mass.  \textsl{Green hatching} and \textsl{red
cross hatching} indicate convective and semiconvective regions.
\textsl{Blue shading} indicates energy generation and \textsl{pink
shading} energy loss.  Both take into account the sum of nuclear and
neutrino loss contributions.  The \textsl{thick black line} at the top
indicates the total mass of the star, being reduce by mass loss due to
stellar winds.  Note that the mass loss rate actually increases at
late times of the stellar evolution.  The decreasing slope of the
total mass of the star in the figure is due to the logarithmic scale
chosen for the time axis.  }
\label{hegerKD}
\end{figure}

As massive stars we denote those that are born with initial masses of
more than about 8\,$\Msun$, the minimum mass for single stars to
explode as supernova.  Once the star has formed, its center generally
evolves to increasing central density and temperature.  This overall
contraction is interrupted by phases of nuclear fusion --- hydrogen to
helium, helium to carbon and oxygen, then carbon, neon, oxygen and
silicon burning, until finally iron is produced and the core
collapses.  Each fuel burns first in the center, then in a shell.  In
Table~\ref{heger:tau} we summarize the burning stages and their
durations for a $20\,\Msun$ star and in Figure~\ref{hegerKD} we show
the evolution of the interior structure of a $22\,\Msun$ star.  The
time scale for helium burning is about ten times shorter than that of
hydrogen burning, mostly because of the lower energy release per unit
mass.  However, the time scale of the burning stages beyond central
helium-burning is radically reduced by thermal neutrino losses that
carry away energy \textit{in situ}, instead of requiring that it be
transported to the stellar surface.  These losses increase with
temperature, roughly $\propto\,T^9$.  (See \cite{WHW02} for a more
extended review.)  When the star has built up a large enough iron
core, exceeding its Chandrasekhar mass, it collapses to form a neutron
star or a black hole.  A supernova explosion may result \cite{FK01},
or even, in rare cases, a gamma-ray burst \cite{MW99,MWH01}.

\begin{table}[b]
\caption{Nuclear burning stages in massive stars.  We give typical
temperatures and time scales for a $20\,\Msun$ star (Pop I; similar in
Pop III) and a $200\,\Msun$ star (Pop III){}\label{heger:tau}}
\vspace{-\baselineskip}
\begin{center}
\setlength\tabcolsep{2ex}
\begin{tabular}{cccccc}
\hline\noalign{\smallskip}
\multicolumn{2}{c}{Burning stages} &
\multicolumn{2}{c}{$20\,\Msun$ star} &
\multicolumn{2}{c}{$200\,\Msun$ star} \\
\noalign{\smallskip}
Fuel & Main product & T ($10^9\,$K) & duration (yr) &  T ($10^9\,$K) & duration (yr) \\
\hline
\noalign{\smallskip}
H  & He     & 0.037 & $8.1\Ex6$ & 0.14           & $2.2\Ex6$ \\
He & O, C   & 0.19  & $1.2\Ex6$ & 0.24           & $2.5\Ex5$ \\
C  & Ne, Mg & 0.87  & $9.8\Ex2$ & 1.1$^\dagger$  & 4.5 \\
Ne & O, Mg  & 1.6   & 0.60      & 2.4$^\dagger$  & $1.1\Ex{-6}$ \\
O  & Si, S  & 2.0   & 1.3       & 3.5$^\dagger$  & $3.5\Ex{-8}$ \\
Si & Fe     & 3.3   & 0.031     & 4.3$^\ddagger$ & $2.7\Ex{-7}$ \\
\noalign{\smallskip}
\hline
\noalign{\smallskip}
\end{tabular}
\end{center}
\vspace{-\baselineskip}
$^\dagger$central radiative implosive burning \hspace{3ex}
$^\ddagger$incomplete silicon burning at bounce
\end{table}

\section{Modern massive stars fates}

\begin{figure}[t]
\begin{center}
\includegraphics[angle=-90,width=\columnwidth,bb=36 126 574 666]{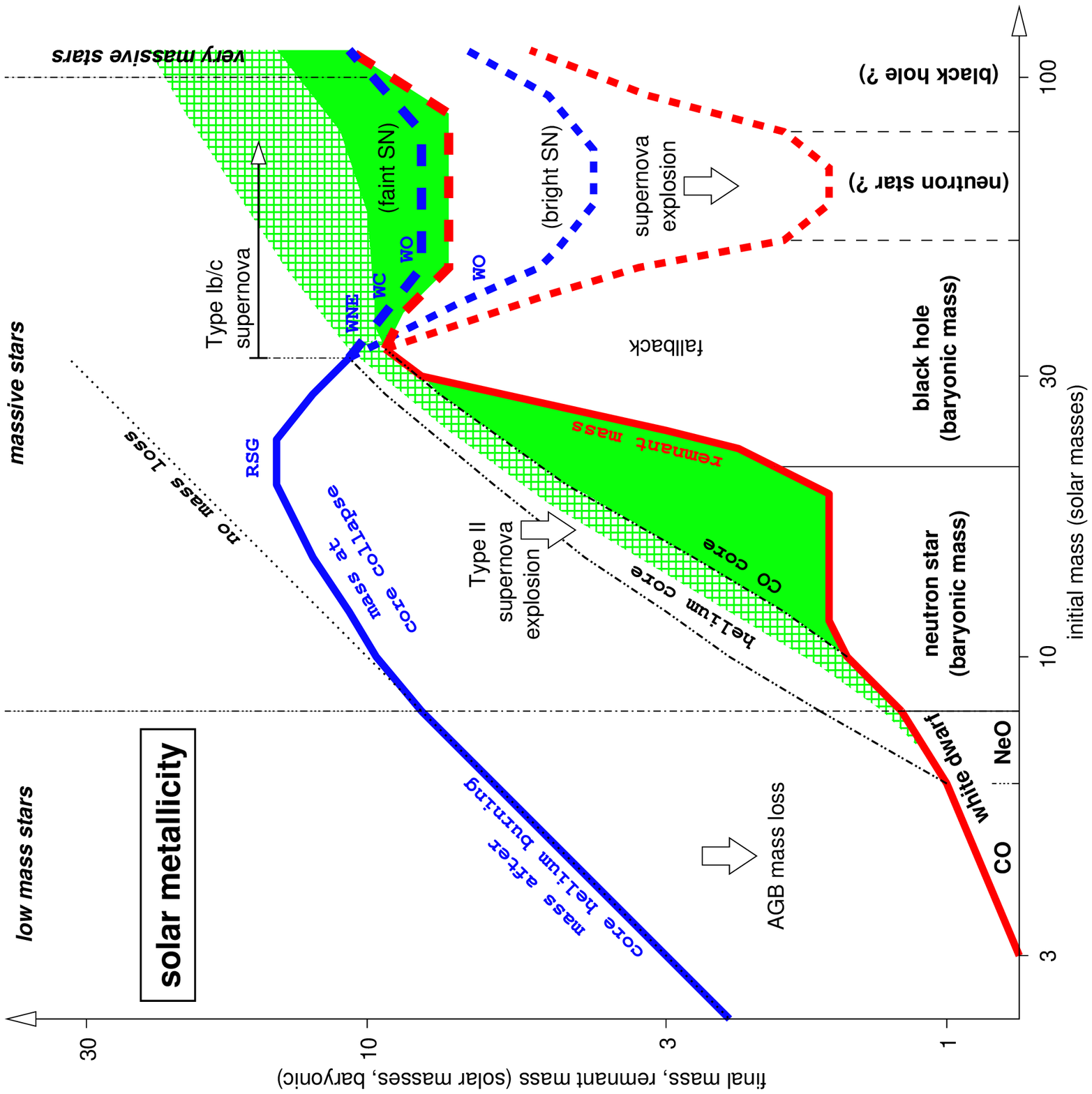}
\end{center}
\caption[]{ Stellar mass at time of final explosion/remnant formation
(\textsl{blue line}), remnant mass (\textsl{red line}) and metals
released (\textsl{green fill} and \textsl{hatching}) as a function of
initial mass of the star for modern stars (Pop I)}
\label{hegerI}
\end{figure}

In Figure~\ref{hegerI} we show the fate of modern stars, formed from a
composition comparable to that of our sun (Pop I), as a function of
initial mass.  Below $\sim8\,\Msun$ initial mass only white dwarfs are
formed, neon-oxygen white dwarfs immediately below this limit, and
carbon-oxygen white dwarfs at even lower masses.  The envelope of such
stars is lost during their AGB stage.

The regime of initial masses above $\sim8\,\Msun$ we refer to as
\emph{massive stars}.  Except for the pair-instability supernovae
discussed below They leave behind compact remnants.  Up to an initial
mass of $\sim20-25\,\Msun$ \cite{Fry99,FK01} typically neutron stars
result.  The layers above the neutron star are ejected, consisting of
the ashes of the preceding hydrostatic stellar burning phases and
explosive burning products (\textsl{green cross hatching} indicates
partial helium burning, i.e., mostly C and O; \textsl{solid green}
indicates pure metals, i.e., products of complete helium burning and
later burning phases).

At higher masses a large fraction of the core can fall back onto the
neutron star after an initial explosion has first driven the matter
outward.  If the resulting remnant mass exceeds the maximum mass for a
neutron star, it collapses to a black hole.  In Figure~\ref{hegerI} we
indicate this region by ``fallback'' and ``black hole''.

The evolution of modern massive stars is significantly affected by mass
loss due to stellar winds.  The mass loss \cite{NJ90} can become so
strong that the final mass of the star may actually decrease as its
initial mass increases.  Indeed, \cite{WHW02} find a maximum in the
final mass around $\sim20\,\Msun$.  In Figure~\ref{hegerI}) the
\textsl{blue curve} shows the final mass at the time of core collapse
for massive stars.  Around $\lesssim35\,\Msun$ the mass loss becomes
so strong that entire hydrogen envelope is lost prior to explosion of
the star: in Figure~\ref{hegerI} the curve for the final mass of the
star intersects the curve for helium core mass at core collapse.  The
exact mass where this happens depends on, e.g., the uncertainty of the
mass loss rates and can vary with initial stellar
rotation \cite{sch92,sch93,mey94,Lan97,HLW00,MM00,MM01}.

Once the massive stars have lost their hydrogen envelope, they become
Wolf-Rayet (WR) stars.  The mass loss of these objects is known to be
large, but also quite uncertain.  Recently, the introduction of
``clumping'' into modeling of Wolf-Rayet star spectra lead to a
reduction of the derived mass loss rates by a factor 2-3 \cite{HK98}.
In Figure~\ref{hegerI} we show for both cases remnant mass and stellar
mass at core collapse: the \textsl{dotted lines} are for the case of
the previously assumed high mass loss rates and the \textsl{dashed
lines} for the case of the lowered Wolf-Rayet mass loss rates.  In the
former case, there may be a regime of initial masses in which the core
mass at the time of explosion is low enough to form neutron stars,
while it is absent in the other case.  The final masses may be higher
than indicated in the figure if the star is covered by a hydrogen
envelope for a significant fraction of central helium burning
\cite{bra01}.

Note that the Wolf-Rayet stars can already release products of central
helium burning by stellar winds before explosion (\textsl{green
hatching} above the \textsl{dashed blue curve} indicating the final
mass of the star) -- or even if they do not explode.  Such objects are
denoted as WC and WO stars.  Here we show the metal production only
for the case of the low WR mass loss rate.

\section{Primordial stars}

\begin{figure}[t]
\begin{center}
\includegraphics[angle=-90,width=\columnwidth,bb=35 120 577 669]{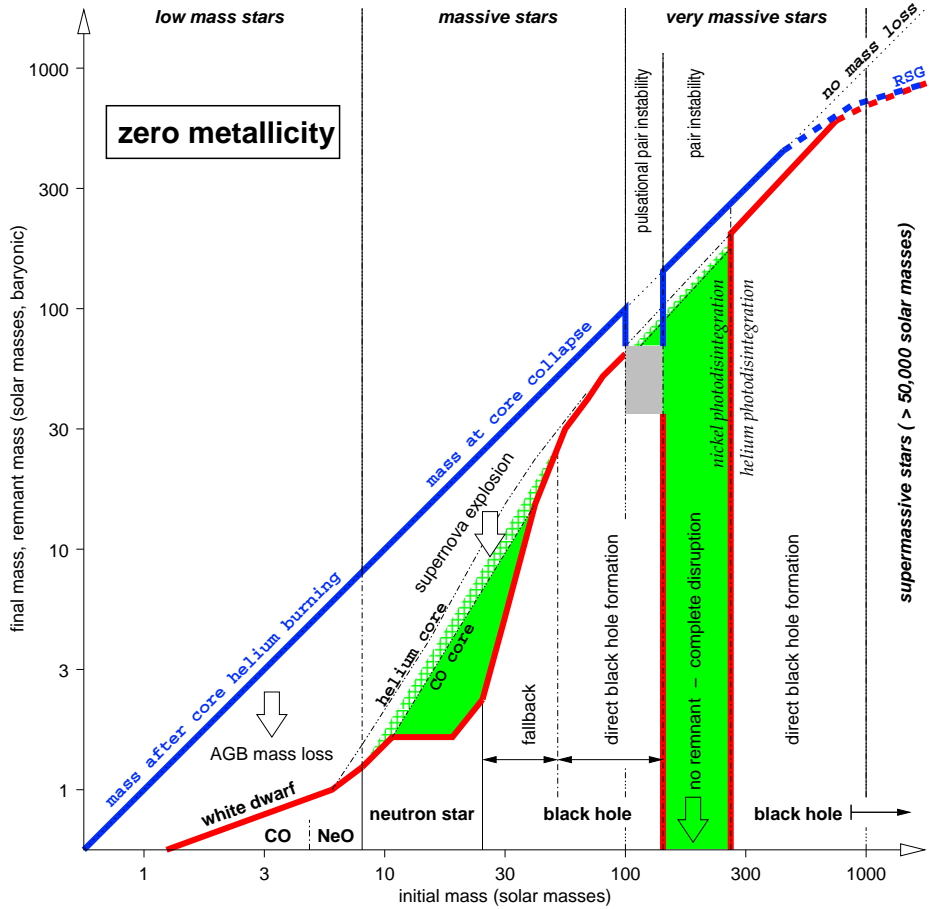}
\end{center}
\caption[]{Stellar mass at time of final explosion/remnant formation
(\textsl{blue line}), remnant mass (\textsl{red line}) and metals
released (\textsl{green fill and hatching}) as a function of initial
mass of the star for primordial (metal-free, Pop III) stars
{}\label{hegerIII}}
\end{figure}

The lack of initial metals in stars that form from the composition as
made in the big bang results in a strongly reduced mass loss, down to
the point of negligibility \cite{Kud02}.  For simplicity, we can
therefore assume that Pop III stars keep almost all of their initial
mass through the end of central helium burning (Figure~\ref{hegerIII};
see also \cite{mar01}).  Recent studies indicate that the first
generation of stars may have been rather massive
\cite{bro01,bro02,ABN02}, or may have had at least a very massive
component \cite{NU00} compared to modern stellar populations.  But
even those stars might keep most of their initial mass till the end of
their evolution \cite{BHW01,HW02}.

For masses below $\lesssim35\,\Msun$ the fate of the stars is similar
to that of modern stars (Figure~\ref{hegerI}) except that the mass at
core collapse continues increasing.  Low-mass stars form white dwarfs,
massive stars first form neutron stars, then black holes by fall back.
However, as the helium core mass at core collapse increases,
eventually a successful supernova shock cannot be launched due to the
strong infall of the increasingly larger oxygen and silicon core
masses.  A black hole is formed directly and no supernova explosion
should occur.

Above an initial mass of $\sim100\,\Msun$, the stars encounter the
electron-positron pair creation instability after central carbon
burning \cite{BAC84,Woo86}.  This leads to rapid burning of oxygen and
silicon (Table~\ref{heger:tau}).  Only when enough energy is released,
the infall is stopped and reverts into an explosion.  Below an initial
mass of $\sim140\,\Msun$ the explosion energy is not big enough to
disrupt the entire star.  The outer layers, the hydrogen envelope and
maybe part of the helium core, however, are ejected.  The rest of the
star falls back and may encounter subsequent pulses of this kind
within one to several 10,000\,yrs until it finally collapses and
directly forms a black hole.  These stars will not have a hydrogen
envelope at the time of collapse, but the ejecta may still be close
by.  The energy of these eruptions can be as high as
$3\times10^{51}\,$erg.

For initial masses from $\sim140-260\,\Msun$ the explosion energy of
the first pulse is already sufficient to entirely disrupt the star.
In this case no remnant remains and all the metals are ejected.  The
explosion energies may reach $\sim10^{53}\,$erg and more than
$50\,\Msun$ of radioactive $^{56}$Ni may be ejected -- both figures
are close to $100\times$ that of Type Ia supernovae.  These events are
rather bright and could be observable out to the edge of the universe
in the infrared \cite{heg02}.  Further observational signatures should
be their long time scale, already in their rest frame, but
additionally boosted by their high redshift, and a Lyman-alpha cut-off
corresponding to their redshift, as they should explode when most of
the universe is not yet re-ionized.

Above initial masses of $\sim260\,\Msun$ photo-disintegration of
alpha-\-particles (which themselves are already the result of
photo-disintegration of iron group elements which were made in silicon
burning) reduces the pressure enough that the collapse of the star is
not turned around but directly continues into a black hole
\cite{Woo86,FWH01}.  Probably no metals are ejected in this case.
Above several $100\,\Msun$ even primordial stars may evolve into red
supergiants, becoming pulsationally unstable and lose mass
\cite{BHW01}.  Since we have no good estimate of the associate mass
loss, we draw the line for the pre-collapse and remnant mass as dashed
lines in this regime.

\section{Supernova and remnant populations}

\begin{figure}[t]
\begin{center}
\includegraphics[angle=-90,width=\columnwidth,bb=35 31 577 760]{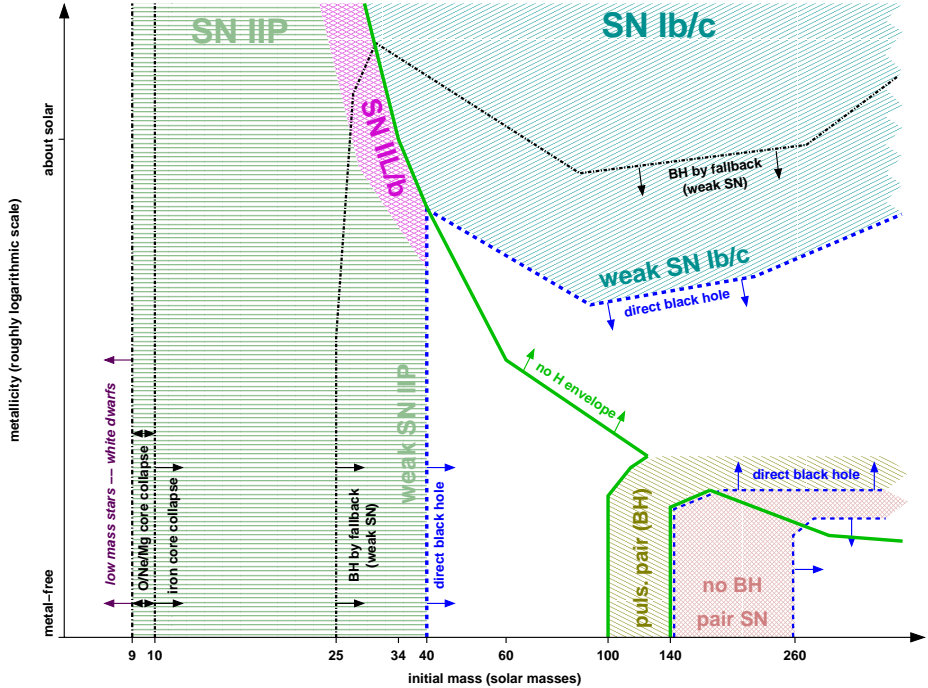}
\end{center}
\caption[]{Supernova types of single stars as a function of stellar
initial mass and metallicity.  \textsl{White regions} show where no
supernova explosion is expected.  \textsl{green horizontal hatching}
indicates Type IIL supernovae, \textsl{purple diagonal hatching} Type
IIb supernovae, \textsl{dark green diagonal hatching} Type Ib/c
supernovae, and \textsl{light green diagonal hatching} and \textsl{red
cross hatching} indicate pulsational and non-pulsational
pair-instability supernovae.  The \textsl{green curve} is the dividing
line between stars that have a hydrogen envelope (below the
\textsl{green line}) and those that don't.  {}\label{hegerSN}}
\end{figure}

\begin{figure}[t]
\begin{center}
\includegraphics[angle=-90,width=\columnwidth,bb=35 31 577 760]{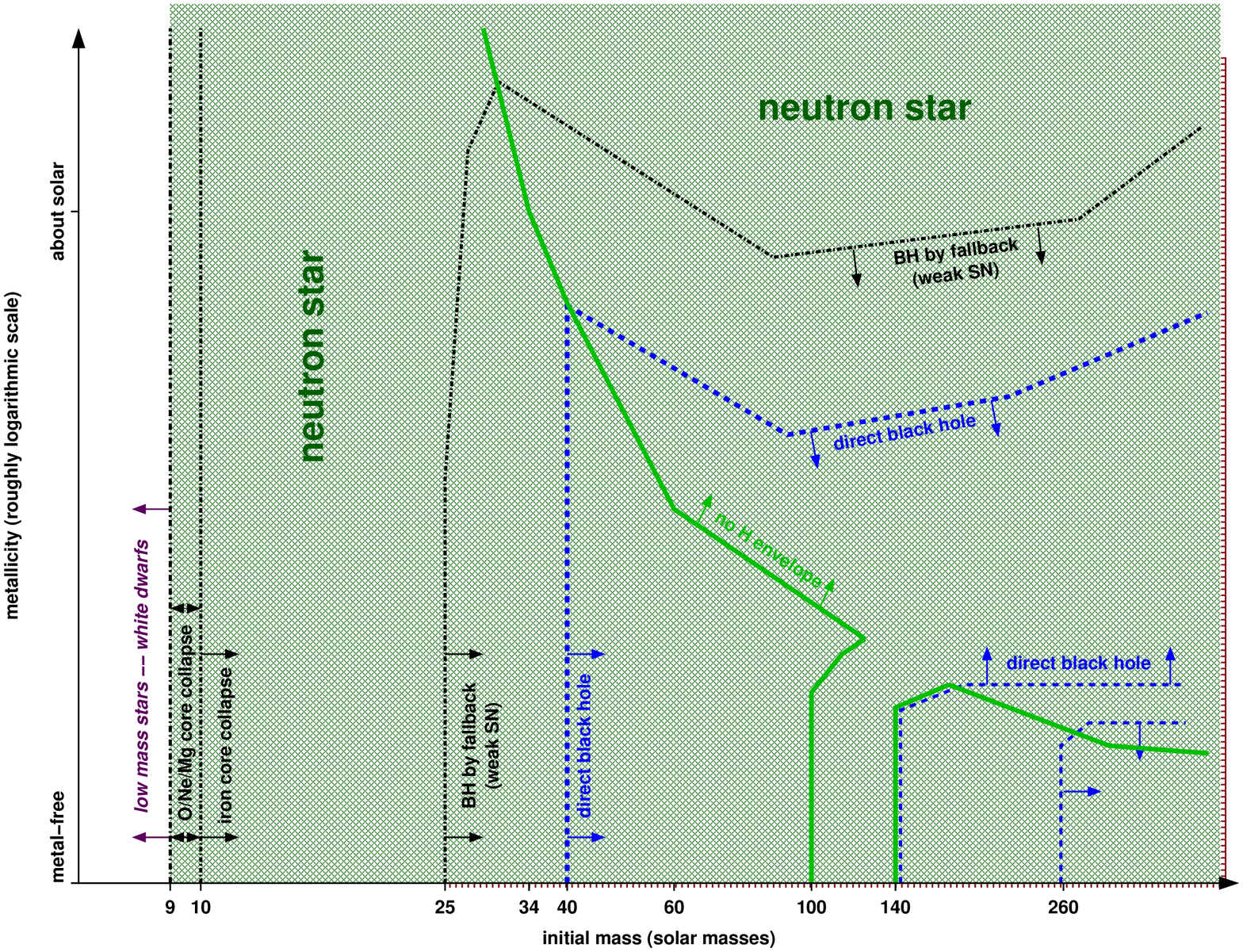}
\end{center}
\caption[]{Stellar remnants of single stars as a function of stellar
initial mass and metallicity.  {}\textsl{Green diagonal cross
hatching} indicates the where neutron stars are made, \textsl{red
horizontal and vertical cross hatching} indicates regimes where black
holes are formed by fall back and \textsl{solid black} shows where
black holes form directly without launching a supernova shock from a
proto-neutron star.  In the \textsl{white region} at low metallicity
and high mass no remnant is left (pair instability supernovae) and in
the \textsl{white region} on the left hand side white dwarfs are made.
We use the same curves and dividing lines as in Figure~\ref{hegerREM}.
{}\label{hegerREM}}
\end{figure}

In Figures~\ref{hegerSN} and \ref{hegerREM} we try to give a rough
sketch of the remnant and supernova types as a function of initial
mass and metallicity.  A very detailed description will be given in
\cite{heg02a}.  Due to the uncertainty of some mass loss rates and
their scaling with metallicity, in particular those of red supergiants
and Wolf-Rayet (WR) stars, we do not give an absolute scale on the
\textsl{y-axis} of these figures, but assume that we can at least
outline the rough sequence of populations with increasing metallicity.

Only at sufficiently low metallicities do very massive stars keep all
their mass till collapse.  As metallicity increases, due to increasing
mass loss, first the very massive stars that directly collapse to
black holes disappear.  Next the pair-instability supernova produce
less and less powerful explosions until they disappear.  Above the
metallicity limit for pair instability all very massive single stars
should leave compact remnants.  There may be a regime in which
pair-instability supernovae can occur in bare helium stars if the
hydrogen envelope has been carried away by stellar wind, but the
helium core has not been shrunk enough by WR winds.  At higher
metallicities also the pulsational pair-instability supernovae will
disappear.  They should always collapse as hydrogen-free stars
(``window'' in the \textsl{green curve} indication the presence/loss
of the hydrogen envelope) and their core collapse produces a black
hole without launching a supernova explosion.  Depending on whether
stellar winds have uncovered the helium core before collapse due high
enough metallicity or just the pair instability pulsations ejected the
hydrogen envelope, the hydrogen may still be in the vicinity of the
star or be blown far away.

If mass loss does not uncover the helium core, above $\sim40\,\Msun$
black holes are formed directly.  In this case we do not expect a
supernova explosion display.  For stars massive enough, the hydrogen
envelope is likely removed by stellar winds, shrinking the stellar
core.  Since the core collapse and explosion are mostly determined by
the core masses, this alters the fate of those stars.  This means that
above a certain metallicity (and as function of their initial mass)
even stars above the direct black hole limit for hydrogen-covered
stars, can avoid directly forming black holes (Figure~\ref{hegerSN})
and may result in hyrogen-free supernova explosions of Type Ib or Ic
\cite{Fil97,Fil02} (Figure~\ref{hegerSN}).

At even higher metallicity, also the limit for black hole formation
due to fallback can be avoided in hydrogen-free stars.  That is, for
high enough metallicity, all stars may result in neutron stars
(Figure~\ref{hegerREM}).  Those supernovae, resulting from smaller
cores, are potentially brighter than those from the more massive
hydrogen-free stars \cite{EW88} (except pair-instability supernovae).

In stars that keep the hydrogen envelope, down to $\sim20-25\,\Msun$
the fallback may be big enough to form black holes \cite{Fry99}, but
yet a Type II supernova \cite{Fil97} may occur.  If the fallback is
excessive, the supernova may be dim.  Below this mass limit neutron
stars are made.  Stars that have lost most of their hydrogen envelope
but about $\lesssim1\,\Msun$ at the time of core collapse may result
in Type IIL/b supernovae, while those that have left more hydrogen
will make Type IIP supernovae (see also \cite{Fil97,Fil02}).
Naturally, in Figure~\ref{hegerSN} the Type IIL/b supernovae thus
occur in a small stripe just below the limit for complete loss of the
hydrogen envelope, as a function of mass and metallicity.  Since this
curve goes above the limit for successful core collapse supernova
explosions at low metallicity, we anticipate that there will be a
lower metallicity limit for Type IIL/b supernovae from single stars.

\section{Summary and discussion}

We outlined, in a rough scheme, the fate of modern and primordial
massive single stars as a function of their initial mass and
metallicity.  The most important difference caused by the variation of
compositions is the strong mass loss in modern stars.  Because of
this, they may lose their hydrogen envelope above a certain initial
mass.  When this happens, the exposed helium star may cause even
stronger mass loss.  Only at very low metallicity can very massive
stars can retain enough mass to become pair-instability supernovae.
On the other hand, certain types of supernovae that require the loss
of most or all of the hydrogen envelope, occur in single stars only
above a certain level of metal enrichment.

The lines between the different regimes in
Figures~\ref{hegerI}--\ref{hegerREM} are only a rough guidance.  Due
to the interaction of different burning shells (Figure~\ref{hegerKD})
the core masses (silicon core, iron core, etc.) do not increase
monotonously as a function of initial mass or metallicity
\cite{WHW02,heg02a}.  Therefore one cannot expect a unique transition
from one regime to the neighboring, but a rather ragged one, with
detached ``islands'' of one regime within the other.

In contrast to single stars, interacting (``close'') binary stars can
lead to the loss of the hydrogen envelope even in metal-free stars.
Therefore, unfortunately, hydrogen-free stars and supernovae can be
present at any metallicity, and thus an observational determination of
the metallicity limit for the loss of the hydrogen envelope from low
metallicity stellar populations is rendered difficult.  However, the
ratio of supernova types may change when single stars start
contributing to Type IIL/b, and Ib/c supernovae.  Depending on the
initial separation and mass ratio of the binary, the mass exchange can
happen at different evolution stages (most prominent: during hydrogen
burning, during the transition from hydrogen burning to helium
burning, and during or at the end of helium burning).  The earlier the
mass transfer happens, the smaller is the final mass of the star.

Finally, the evidence for an association of gamma-ray bursts (GRBs)
with massive star populations seems to increase \cite{kip98}.
Provided a star that collapses to a black hole has sufficient angular
momentum, it may launch a jet \cite{Woo93}, blowing up the star (e.g.,
\cite{MWH01}; ``hypernova'', see also \cite{nom02}) and possibly
producing a GRB if the jet escapes the star
\cite{Woo93,MW99,ZWM02,ZW02}.  If any single star can ever have
sufficient angular momentum at core collapse, in Figure~\ref{hegerREM}
only in the ``black hole by fallback'' and ``direct black hole''
regimes such events can occur.  And only above the line for loss of
the hydrogen envelope, GRBs could be made.

\vspace{0.5\baselineskip} {\small\textbf{Acknowledgments.}  This
research has been supported by the NSF (AST 02-06111), the DOE ASCI
Program (B347885) and the SciDAC Program of the DOE
(DE-FC02-01ER41176).  AH is supported in part by the Department of
Energy under grant B341495 to the Center for Astrophysical
Thermonuclear Flashes at the University of Chicago and acknowledges
travel support from MPA to participate in this meeting.}

%

\end{document}